%% file: 00_main.tex
\documentclass[11pt]{article}
\usepackage{graphicx} 
\usepackage{biblatex}
\usepackage{amsmath,amssymb}
\usepackage[margin=1in]{geometry}
\usepackage{cleveref}
\usepackage{xcolor}
\usepackage{comment}
\usepackage{authblk}

\title{Characterizing Atomistic Transitions Using Cross-scale Graph-pooled Chebyshev Signatures}
\author[1]{Rostyslav Hnatyshyn}
\author[2]{Danny Perez}
\affil[1]{School of Computing and Augmented Intelligence, Arizona State University}
\affil[2]{X-Computational Physics AI4ND, Los Alamos National Laboratory}
\date{December 2025}

\bibliography{references}
\begin{document}
\maketitle
\begin{abstract}
Large-scale atomistic simulations can produce extreme volumes of information in the form of long trajectories. Reliably and automatically extracting key information from such datasets remains a formidable challenge, especially as it pertains to the analysis of the structural transitions affecting the system. We present a novel approach to characterize and compare atomistic transitions using cross-scale graph-pooled Chebyshev signatures. These signatures are permutation invariants of an operator that transform a Coulomb matrix representation of the initial state of the system into that corresponding to the final state. Using a long-time trajectory of a small metallic nanoparticle, we show that these signatures can be used to define a natural distance metric between transitions that allows for classification and clustering into physically meaningful families. This approach is shown to capture complex patterns and hierarchies of transition types that are inaccessible to traditional techniques, dramatically facilitating the analysis of large-scale simulations. 
\end{abstract}

\section{Introduction}

Atomistic simulations track the evolution of materials and molecules as they undergo rich sequences of structural transformations.
Individual simulations can contain millions of such structural transitions, especially when generated with long-time methods such as Accelerated Molecular Dynamics \cite{perez2009accelerated} or kinetic Monte Carlo \cite{voter2007introduction}. 
The interpretation of these simulations presents a considerable challenge, as it involves the identification and isolation of key structural events and their classification into meaningful families. 
This analysis currently overwhelmingly relies on manual visualization and analysis by domain experts. 
Automating simulation post-processing therefore has the potential to significantly increase the amount of information that can be extracted from computationally intensive simulations and to consequently enable the development of powerful machine learning approaches.

Conventional approaches to trajectory analysis cluster individual frames of a simulation in an appropriate structural space (e.g., Cartesian or internal coordinates, RMSD matrices, or more sophisticated descriptors) and interpret clusters as metastable conformations or structural groups. 
In materials simulations, order parameters such as local bond environments or other structural fingerprints (Steinhardt order parameters \cite{steinhardt1983bond}, common neighbor analysis \cite{radhi2017identification}) are commonly used to differentiate crystalline phases, amorphous regions, or defect types. 
These descriptors can be used to track the structural evolution of the material along the simulation trajectory and hence to robustly identify different regimes or trends. 
These conventional state-based approaches focus on the sequence of configurations that are visited, but not on the process by which individual states transform into each other.  

Kinetic frameworks instead construct coarse-grained models that can be used to cluster states based on dynamical consideration. 
For example, Markov state models \cite{pande2010everything,bowman2013introduction} and related transfer-operator or Koopman approaches \cite{wu2017variational} estimate a global propagator over a configuration or feature space. 
Spectral analysis of these operators, through eigenvalues and eigenfunctions or through methods such as Perron cluster cluster analysis \cite{deuflhard2005robust}, can then be used to identify metastable macrostates and slow dynamical modes that capture key kinetic events. 
However, these methods do not intrinsically enable the comparison of the different transition pathways observed in different stages of a simulations.

A newer generation of methods pushes further toward a transition-centric view by directly clustering trajectory segments or paths. 
CATBOSS \cite{damjanovic2021catboss}, for example, combines change-point detection with density-peak clustering to classify short trajectory segments into metastable pieces and brief transition intervals, thereby explicitly labeling rare events in time.
Trajectory map visualizations \cite{kovzic2024trajectory} use both structural similarity and temporal ordering to highlight interstate transitions and their relative rarity. 
In biomolecular applications, Path Similarity Analysis \cite{seyler2015path} quantifies distances between entire high-dimensional transition paths and clusters them to reveal distinct mechanistic routes between the same end states. 
Together, these developments underscore a growing recognition that complex systems evolve through ensembles of transitions pathways whose diversity and organization are scientifically important.

Despite this progress, there remains a noticeable gap for systems --- particularly hard materials --- in which simulations generate very large, heterogeneous populations of relatively local events: vacancy or interstitial hops in different environments, slips on various planes, local bond-rearrangement motifs during amorphization or recrystallization, or shape fluctuations in metallic nanoparticles. 
Indeed, even apparently simple systems are known to evolve through large sets of collective multi-atoms reaction pathways \cite{henkelman2003multiple} that can occur in rapid succession. 
In such cases, the objects of interest are the individual elementary transitions that represent the system before and after structural changes localized in space and time. 
Existing tools provide limited support for automatically \textit{classifying and organizing such events at scale}. 
Frame clustering tells us which states are visited, but only through manual inspection can one identify recurring the local rearrangements mediating those transitions. 
Global kinetic models reveal which state-to-state jumps are slow or frequent, but they do not, by themselves, compare or classify the mechanisms that connect different sets of metastable states. 
Segment- and path-based approaches are powerful when one has a modest number of well-resolved trajectories, but they become cumbersome when confronted with millions of short, heterogeneous events distributed across large systems.

There is therefore a need for methods that treat transitions in metastable systems as first-class objects. 
In this work, we take steps in that direction by showing how transitions can be labeled and compared in a way that (i) captures how the underlying topology changes between initial and final configurations, (ii) is agnostic to where in the system the event occurs, and (iii) can be readily analyzed using unsupervised learning approaches to discover families of recurrent transition motifs. 
Our approach enables the automatic construction of ``transition taxonomies'' for complex MD datasets: organizing local events into mechanistic families, quantifying their relative frequencies, and relating them to macroscopic observables.

The paper is organized as follows. 
In \Cref{sec:methods}, we formally define transition operators between representations of the physical system before and after a transition and propose permutation-invariant signatures that can be used to characterize and compare transitions, which we refer to as cross-scale Chebyshev signatures. 
In \Cref{sec:results}, we then illustrate how these signatures can be used to automate the analysis of very large simulations of metallic nanoparticles that evolve through an extremely rich set of complex collective transitions. We finally discuss the properties of this method in comparison with traditional approaches in \Cref{sec:discussion} before concluding.

\include{01_methods_cross_scale_energy}

\include{02_results}
\include{03_discussion}


\printbibliography

\end{document}

%% file: 01_methods_cross_scale_energy.tex
\section{Methods}
\label{sec:methods}

In the following, we construct a \textit{transition operator} that mathematically describe the changes between the representations of a complex atomistic system before and after a structural transition. We then show how invariants of this operator can be used as signatures in a way that enables meaningful labeling and pairwise comparisons.

\subsection{Problem setting}
We consider pairs of atomic configurations $\mathbf{R}_0,\mathbf{R}_1 \in \mathbb{R}^{N\times 3}$ containing the same number of atoms $N$. Each element of a pair corresponds to the initial and final states of a structural transition $\mathbf{R}_0 \rightarrow \mathbf{R}_1$. In the following, we consider representations of these atomic configurations in terms of matrices  $\mathbf{C}_0(\mathbf{R}_0),\mathbf{C}_1(\mathbf{R}_1) \in \mathbb{R}^{N\times N}$.
The choice of the mapping between $\mathbf{R}$ and $\mathbf{C}$ is not unique; we here directly draw from the concept of Coulomb matrices \cite{rupp2012fast} to define each element $i,j$ of $\mathbf{C}$ in terms of  functions of the Cartesian distance between atoms $i$ and $j$. Throughout, we assume that pairs of $\mathbf{R}_0,\mathbf{R}_1$ and $\mathbf{C}_0,\mathbf{C}_1$ corresponding to a given transition are consistently ordered, i.e., that corresponding rows of these objects refer to the same physical atom. We also assume that the $\mathbf{C}$ matrices are symmetric positive definite (SPD), which simplifies the derivation and numerical implementation.   To enforce that $\mathbf{C}$ remains SPD, we employ Wendland kernels \cite{wendland1995piecewise}, a family of finite-range piecewise polynomials that yield provably SPD matrices for any non-pathological atomic configuration.

For each transition, we then define the matrix $\mathbf{T}  \in \mathbb{R}^{N\times N}$ as the unique SPD solution of the congruence equation:
\begin{equation}
\mathbf{T}\,\mathbf{C}_0\,\mathbf{T} = \mathbf{C}_1.
\end{equation}
This equation is trivially solved by
\begin{equation}
\mathbf{T} = \mathbf{C}_0^{-1/2}\left(\mathbf{C}_0^{1/2} \mathbf{C}_1 \mathbf{C}_0^{1/2}\right)^{1/2}\mathbf{C}_0^{-1/2}.
\end{equation}
The matrix $\mathbf{T}$ can be thought of as a discrete operator that transforms the representation $\mathbf{C}_0$ of the initial configuration of the atomistic system into that of the final configuration $\mathbf{C}_1$. $\mathbf{T}$ hence contains sufficient information to completely represent the geometric changes that occur during a transition (given that the $\mathbf{C}$ matrices defined here can be used to reconstruct atomic positions up to rigid translations and rotations when the truncation radius $r_c$ of the Wendland kernels is sufficiently large). Note that because the $\mathbf{C}$ and $\mathbf{T}$ matrices are SPD, matrix functions such as the square root can easily be applied through the conventional spectral approach. 

The fact that many transitions in extended systems are spatially localized is reflected in the structure of the matrix $\mathbf{T}$. For example, spatial regions that are undisturbed by a local transition correspond into a block of $\mathbf{T}$ that approaches identity, or equivalently, by eigenvectors of $\mathbf{T}$ corresponding to eigenvalues that approach 1. In general, it is desirable to nullify the impact of these trivial ``inert" regions when manipulating transition matrices so that transitions occurring in different systems can be compared. To do so, we introduce a new matrix $\mathbf{B}$ through the relation
\begin{equation}
\mathbf{B} = \log \mathbf{T},
\end{equation}
where the log is the matrix operation, not an element-wise application. In the spectral representation of matrix functions, this transformation zeroes out the trivial ``inert" eigenvalues, and hence the contribution of the corresponding eigenvectors. This makes $\mathbf{B}$ even sparser and localized. 
This matrix, which contains the same information as the original matrix $\mathbf{T}$, is the central object of the proposed approach. 

While the matrices $\mathbf{B}$, $\mathbf{C}$, and $\mathbf{T}$ are rotation and translation invariant by the virtue of being expressed only in terms of pairwise distances, their element-wise structure depends on an  arbitrary numbering convention, so that distances between two transition matrices $\mathbf{B}$ and $\mathbf{B'}$ cannot simply be measured through a matrix norm of the form $|\mathbf{B}-\mathbf{B'}|$.  Instead, we seek a representation \(\Phi(\cdot)\) and induced distance \(d(\mathbf{B} ,\mathbf{B'} )\) that are invariant to \emph{independent} permutations/renumbering operations $P$ and $Q$, i.e.,:
\begin{equation}
d(\mathbf{B},\mathbf{B'})=d(\mathbf{P}^\top \mathbf{B} \mathbf{P},\; \mathbf{Q}^\top \mathbf{B} \mathbf{Q})~ \forall~\mathbf{P},\mathbf{Q}
\end{equation}
This property is required so that topologically identical transitions occurring in different regions of the system can be be recognized as such, i.e., we are interested in characterizing the distance between topologically-equivalent transition classes, not between particular instances of these transitions.

In the following, we develop an approach to define a \emph{permutation-invariant} signature for transition matrices $\mathbf{B}$ that will allow for a natural definition of a distance between transitions. In general, it is thought to be impossible to design a complete permutation-invariant signature of $\mathbf{B}$, as this would allow for a general solution to the graph isomorphism problem. Any signature can therefore be expected to be "lossy". E.g., a natural choice for such a signature is the sorted spectrum of $\mathbf{B}$, an approach that was leveraged in the context of Coulomb matrices \cite{rupp2012fast}. However, many different SPD matrices share the same spectrum, as was recognized in the context of molecular representations using Coulomb matrices \cite{schrier2020can}.
Intuitively, eigenvalues capture the magnitude of the different components of the transformation, but not their spatial characteristics, which are encoded in their eigenvectors. 
It is therefore advisable to enrich spectral information with information on the spatial correlations between the different components of the transition. 
While eigenvector invariants could directly capture this information (e.g., sorted lists of eigenvector components), we propose a physically grounded representation that systematically capture spatial correlations. 
To do so, we define physically motivated "probe" functions and matrix transformations to jointly capture a combination of the spectral and spatial features of $\mathbf{B}$. 

\subsection{Geometry graph and Laplacian}
\begin{figure}
    \centering
    \includegraphics[width=\linewidth]{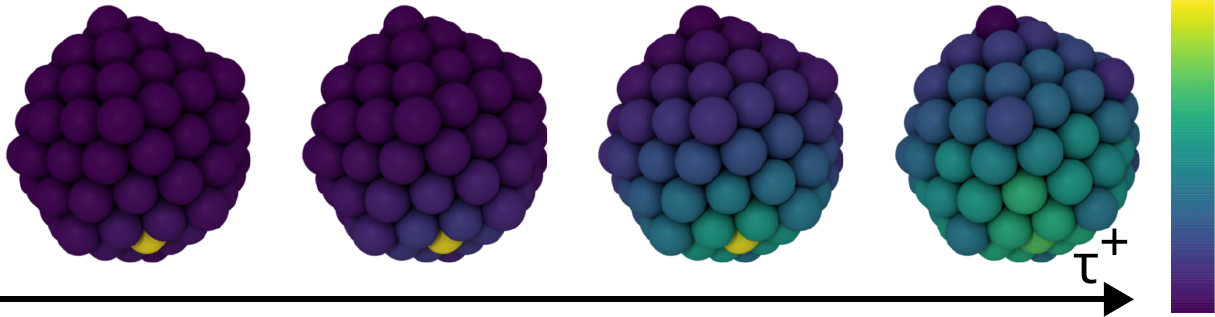}
    \caption{A visualization of the graph smoothing operator $\mathbf{G}$ operating on a single atom within a configuration at different values of $\tau =0,1,5,10$. Darker colors indicate low values, while brighter colors indicate high values; this convention is followed throughout the manuscript. Short time operators are very local probes of spatial correlations, while longer times capture larger structures.}
    \label{fig:g_visualization}
\end{figure}

Spatial information on the initial state of the transition is encoded in the matrix $\mathbf{C}_0$. 
This matrix can be viewed as a weighted undirected graph that encodes the "strength" of the interaction between atoms. 
This forms the canvas on which the transition operates, and is hence a natural tool to characterize it. 
Let \(\mathbf{D}=\mathrm{diag}(d_1,\dots,d_n)\) with \(d_i=\sum_j \mathbf{C_0}_{ij}\); we then compute the conventional normalized graph Laplacian operator \(\mathbf{L}\):
\begin{equation}
\mathbf{L}  = I - \mathbf{D}^{-1/2} \mathbf{C}_0 \mathbf{D}^{-1/2}.
\end{equation}
which is positive semi-definite (PSD). 
We then define a family of symmetric graph smoothing (pooling) operators \(\mathbf{G}_u\) via the graph heat kernel:
\begin{equation}
\mathbf{G}_u \;:=\; \exp(-\tau_u \mathbf{L}),\qquad u=1,\dots,U
\end{equation}
where \(\tau_u>0\) are scale parameters.
Its action on a basis vector \(e_i\) produces a localized averaging kernel centered at node \(i\), whose effective support increases with \(\tau_u\). This set of pooling operators is therefore an ideal tool to spatially probe the eigenvectors of transition matrices at different locations and scales. 
\Cref{fig:g_visualization} provides an example of the operator centered on an arbitrary atom at different values of $\tau$ illustrating the diffusive analogy where the probe function is extremely localized at short times and becomes diffuse at longer times.

\subsection{Spectral filtering using Chebyshev polynomials}
Following a similar approach, we also expand the spectral signature space by considering various matrix functions of $\mathbf{B}$ through an expansion in terms of Chebyshev polynomials. 

To obtain stable polynomial matrix functions, we first scale \(\mathbf{B}\) so that its eigenvalues lie in \([-1,1]\). Let \(\lambda_{\min},\lambda_{\max}\) denote bounds on the smallest and largest eigenvalues of $\mathbf{B}$. We define the affine spectral map
\begin{equation}
\widetilde{\mathbf{B}} \;=\; \frac{2\mathbf{B}-(\lambda_{\max}+\lambda_{\min})\mathbf{I}}{\lambda_{\max}-\lambda_{\min}},
\end{equation}
which ensures \(\mathrm{spec}(\widetilde{\mathbf{B}})\subseteq [-1,1]\). When comparing different transitions, all corresponding matrices $\mathbf{B}$ are normalized using the \textit{same} transformation that simultaneously maps all of their spectra to the interval $[-1,1]$. To keep the notation light, we will assume in the following that all $\mathbf{B}$ matrices are already transformed accordingly.

\begin{figure}
    \centering
    \includegraphics[width=\linewidth]{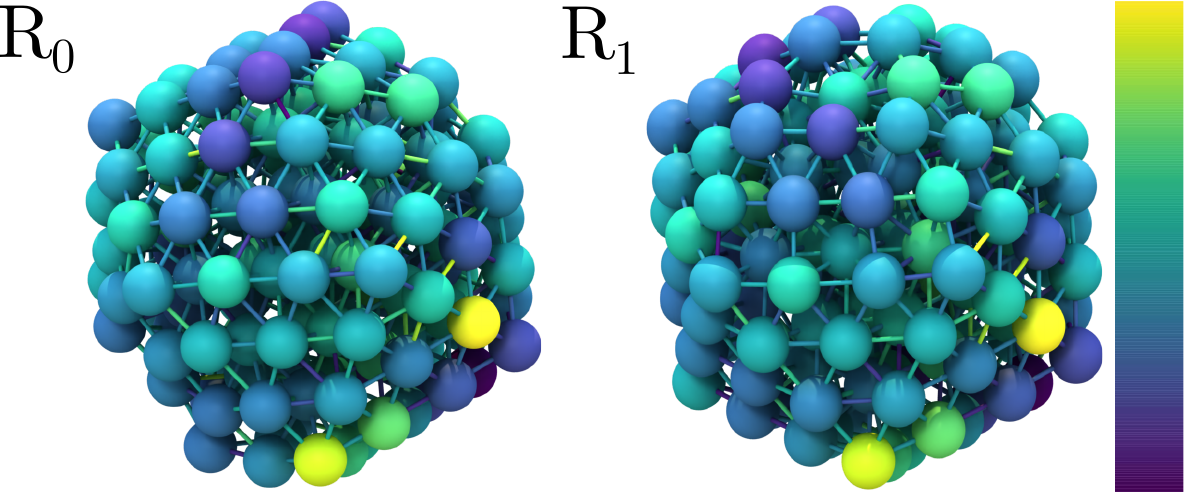}
    \caption{Absolute values of $\mathbf{A}_1$ for an arbitrary transition. Atoms are colored by diagonal entries and bonds by the off-diagonal entries.}
    \label{fig:a_visualization}
\end{figure}

With the spectrum normalized, let \(T_k(\cdot)\) be the Chebyshev polynomials of the first kind, defined by \(T_0(x)=1\), \(T_1(x)=x\), and the recurrence
\begin{equation}
T_{k+1}(x) = 2x\,T_k(x) - T_{k-1}(x).
\end{equation}
We define Chebyshev matrix polynomials of \( \mathbf{B}\) by
\begin{equation}
\mathbf{A}_k \;:=\; T_k(\mathbf{B}),\qquad k=0,1,\dots,K.
\end{equation}
These matrices are computed using the corresponding matrix recurrence:
\begin{align}
\mathbf{A}_0 &= I,\\
\mathbf{A}_1 &= \mathbf{B},\\
\mathbf{A}_{k+1} &= 2\mathbf{B} \mathbf{A}_k - \mathbf{A}_{k-1}\quad (k\ge 1).
\end{align}
Interpreting this expansion through the lens of spectral matrix functions corresponds to transforming the spectrum through different polynomial filters, which has the effect of enhancing certain sections of the spectrum and inhibiting others. 
The different $\mathbf{A}_{k+1}$ matrices therefore offer different spectral windows into the transition operator; \Cref{fig:a_visualization} provides an illustrative example for $\mathbf{A}_1$.

\subsection{Invariant signature definition}
We combine spectral filtering with geometric pooling by defining, for each Chebyshev order \(k\) and scale pair \((u,v)\):
\begin{equation}
\mathbf{S}_{u,v,k} \;:=\; \mathbf{G}_u\,\mathbf{A}_k\,\mathbf{G}_v.
\end{equation}
This operator may be interpreted as a cross-scale coupling: \(\mathbf{G}_u\) aggregates information over different spatial neighborhoods at scale \(\tau_u\), \(\mathbf{A}_k\) applies a spectral filter of \(\mathbf{B}\), and \(\mathbf{G}_v\) re-aggregates outputs at scale \(\tau_v\). When \(u\neq v\), \(\mathbf{S}_{u,v,k}\) is generally not symmetric.

To robustly include spatial information, we summarize each \(S_{u,v,k}\) by its node-wise squared row norms:
\begin{equation}
e^{(u,v)}_{i,k} \;:=\; \left(\mathbf{S}_{u,v,k}\mathbf{S}_{u,v,k}^\top\right)_{ii}
\;=\; \sum_{j=1}^n \left(\mathbf{S}_{u,v,k}\right)_{ij}^2
\;=\; \big\|\left(\mathbf{S}_{u,v,k}\right)_{i,:}\big\|_2^2.
\end{equation}
Equivalently, in quadratic-form notation:
\begin{equation}
e^{(u,v)}_{i,k} = e_i^\top\,\mathbf{G}_u \mathbf{A}_k\,\mathbf{G}_v^2\,\mathbf{A}_k\,\mathbf{G}_u\,e_i,
\end{equation}
which demonstrates that these energies $e^{(u,v)}_{i,k}$ are diagonals of an SPD matrix and therefore nonnegative. Intuitively, \(e^{(u,v)}_{i,k}\) measures how strongly the scale-\(u\) neighborhood around node \(i\) couples (through the \(k\)-th spectral component of \(B\)) to the graph as seen at scale \(v\).

As node relabeling can permute the indices \(i\),   we form a permutation-invariant multiset feature for each triple \((u,v,k)\) by sorting over all atoms:
\begin{equation}
\phi_{u,v,k}(\mathbf{B}) \;:=\; \mathrm{sort}\left( e^{(u,v)}_{1,k},\dots,e^{(u,v)}_{n,k}\right)\in\mathbb{R}^n.
\end{equation}
The full signature is the concatenation over selected orders and scale pairs:
\begin{equation}
\Phi(\mathbf{B}) \;=\; \mathrm{concat}\Big(\{\phi_{u,v,k}(\mathbf{B})\}_{k=0}^{K}\, \Big),
\label{eq:signatures}
\end{equation}
By construction, \(\Phi(\mathbf{P}^\top \mathbf{B} \mathbf{P})=\Phi(\mathbf{B})\) provided the geometry graphs are permuted consistently as \(\mathbf{L}\mapsto \mathbf{P}^\top \mathbf{L} \mathbf{P}\). Since \(\Phi\) is computed separately for each matrix, any distance applied to signatures is invariant to independent permutations of inputs:
\begin{equation}
d(\mathbf{B},\mathbf{B'})=\mathrm{dist}(\Phi(\mathbf{B}),\Phi(\mathbf{B'}))
\quad\Rightarrow\quad
d(\mathbf{P}^\top \mathbf{B} \mathbf{P},\;\mathbf{Q}^\top \mathbf{B'} \mathbf{Q})=d(\mathbf{B},\mathbf{B'}).
\end{equation}

Finally, we define a distance/similarity measure between pairs of transitions by summing over 1D Wasserstein distances on the sorted signature vectors \(\phi_{u,v,k}\) over the different blocks of the signature. We will demonstrate in \Cref{sec:results} that this general signature efficiently captures topological similarity between highly complex collective transitions.

\subsection{Spatial locality of the signatures}
As discussed above, although it is not the case for the system analyzed here, many transitions in hard materials tend to be spatially localized. 
It is therefore desirable for the signature to reflect this locality so that identical transitions embedded in different environments can still be resolved as so. 
Because $\mathbf{B}$ is sparse and localized over the transition region, the signatures defined above are also local in addition to being permutation invariant. 
However, the repeated products of $\mathbf{B}$ induced through the formation of Chebyshev polynomials $\mathbf{A}_k$ will gradually extend the spatial extension probed by the signatures. Therefore, if strong locality is critical for a particular application, the spectral filtering can be omitted or limited to low orders. 

\subsection{Hyper-parameters}
Our method introduces several hyper-parameters when computing the cross-scale Chebyshev signatures. $r_c$ determines the maximum distance between two interacting atoms in the initial $\mathbf{C}$ matrices, while $k_c$ determines the number of Chebyshev matrices computed for each signature. $t$ denotes the sequence of $\tau$ values used in the calculation of $\mathbf{B}$. Finally, the Wendland kernel function can be selected according to varying degrees of smoothness (see \cite{wendland1995piecewise}).

%% file: 02_results.tex
\section{Results}
\label{sec:results}

In this section, we demonstrate the proposed method on very long-time atomistic simulations generated with the Parallel Trajectory Splicing (ParSplice)  \cite{perez2016long} accelerated molecular dynamics method \cite{perez2009accelerated, perez2024recent}.
ParSplice is designed to increase the timescale amenable to direct simulation for systems that evolve through sequences of quasi-discrete activated events.
It does so using a parallel-in-time approach where trajectory segments are generated in parallel and then assembled into statistically correct long-time trajectories. 
Each metastable state in ParSplice is defined as the basin of attraction of a given local minimum on the high-dimensional potential energy surface. 
These minima are labeled using a canonical graph labeling approach \cite{beland2011kinetic} for bookkeeping purposes. 
In contrast to conventional molecular dynamics approaches, ParSplice does not output frames at fixed time intervals, but instead saves the sequence of minima corresponding to states visited by the trajectory together with the residence time in each state. 
In the following, all analyses are applied to these energy-minimized structures.

We consider the trajectory of a small Platinum nanoparticle containing 147 atoms simulated in the canonical NVT ensemble at 700K, following the simulation protocol reported in  \cite{huang2018direct}. 
Such simulations were previously used as prototypes for investigating slow conformational fluctuations in nanoparticles, which are known to be extremely complex \cite{iijima1986structural,marks1994experimental,ben1997correlated,smith1986dynamic}. 
In spite (or perhaps because) of their small sizes, these systems are extraordinarily rich, as their thermodynamic ground states strongly depend on size and temperature. 
As such, small particles tend to spontaneously transition between bulk-like face-centered cubic (FCC) structures and decahedral or icosahedral structures through extremely complex multi-step pathways \cite{huang2018direct} that encompass many intermediate states. 
Furthermore, collective surface and bulk transitions that involve large numbers of atoms are frequently observed. 
This stands in contrast with defective bulk materials where transitions tend to be strongly localized.
This physical complexity coupled with the extremely large amount of information produced by even a single simulation makes this system a challenging testbed for trajectory analysis \cite{huang2017cluster} and visualization \cite{hnatyshyn2023molsieve,hnatyshyn2026lamda}.

The simulation consisted of about 6,700,000 state-to-state transitions spanning approximately 63 microseconds. 
Using the graph labeling techniques implemented in ParSplice, we determined that approximately 40,000 of these transitions connect unique pairs of states, demonstrating that the trajectory is extremely repetitive; it revisits the same groups of states (which we refer to as super-states) multiple times for long periods of time before the trajectory moves on to new ones. 
This quasi-repetitive and hierarchical structure makes analysis particularly challenging as transitions are far too numerous to inspect and classify manually.
Furthermore, this monumental task would be made largely futile by the fact that most transitions are generally ``uninteresting'' as they correspond to small transient structural changes that are likely to be mostly undone by subsequent transitions. 
In contrast, as will be shown below, only a small number of transitions can be expected trigger persistent topological changes corresponding to a jump to a new super-state. 

\begin{figure}
    \centering
    \includegraphics[width=\linewidth]{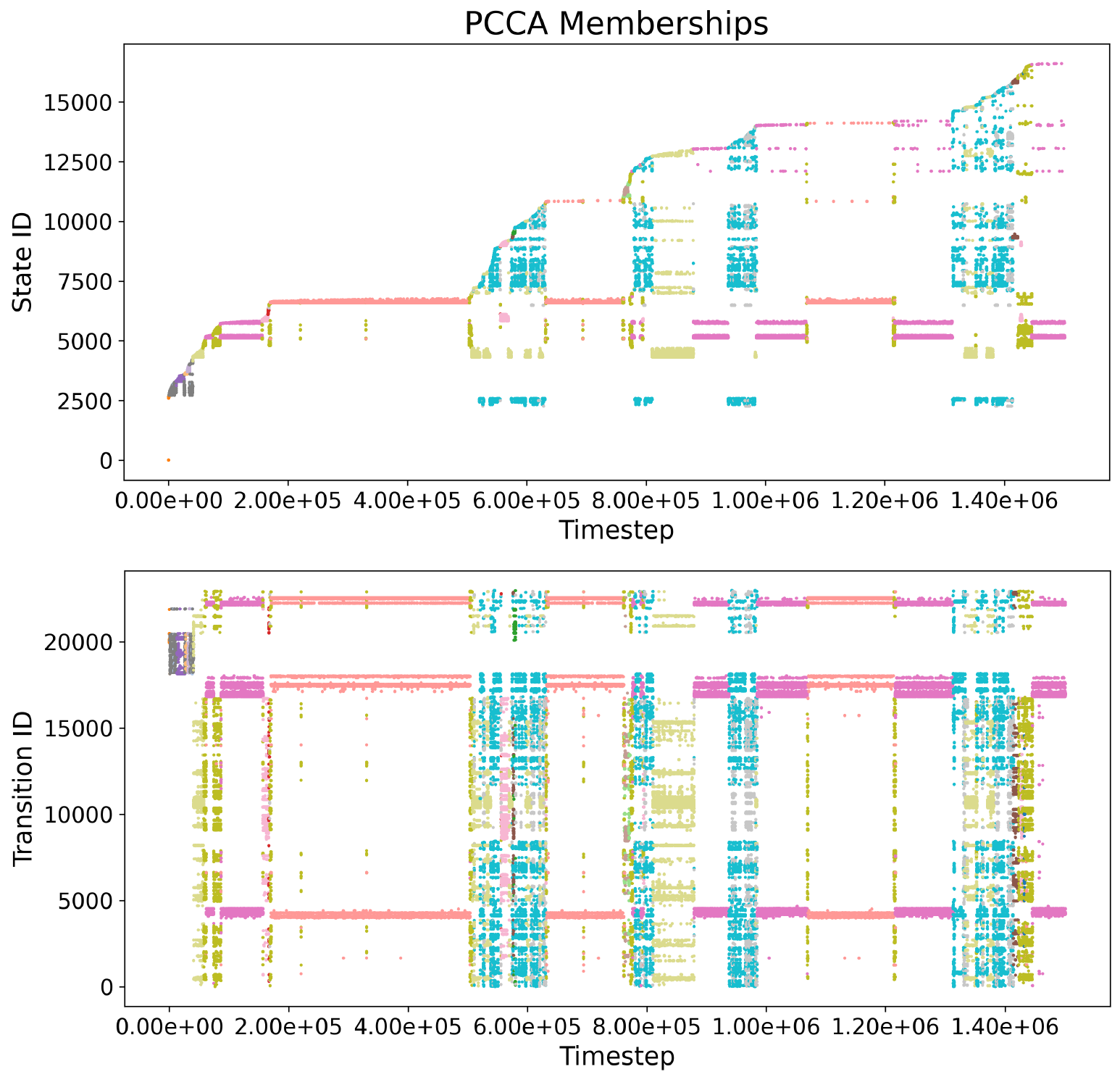}
    \caption{Illustration of the target trajectory. Top: Transition IDs are plotted as a function of the transition number; the transition IDs were reordered based on the leaf ordering of the dendrogram obtained from average linkage hierarchical clustering. Bottom: State ID vs transition number. States and transitions are colored according to their primary PCCA cluster, as identified by the GPCCA algorithm with $K_\mathrm{PCCA}=19$ super-states.  }
    \label{fig:pcca_vs_time}
\end{figure}

We have previously shown \cite{Huang.2017.CAA} that the potentially interesting ``transitional" regions between super-basins can be narrowed down using dynamics-aware clustering techniques such as the Generalized Perron-Cluster-Cluster Analysis~\cite{Reuter19} (GPCCA). 
GPCCA relies on a spectral analysis of an inferred state-to-state transition matrix to identify metastable super-states. 
Because this assignment relies strictly on kinetic information, it is an example of the second class of methods discussed in the introduction. 
GPCCA learns fuzzy membership probabilities that assigns each frame to a ``macro" super-state: a probability close to 1 on a particular super-state indicating a strong assignment while "mixed" probabilities or rapid temporal changes in super-state assignment potentially indicate transitional regions. These transitional regions were shown to correspond to periods of time where long-lived changes in the topology of the nanoparticle tend to occur \cite{Huang.2017.CAA}.   

\Cref{fig:pcca_vs_time} reports a summary of the trajectory that will be analyzed below in the form of the unique discrete ID assigned to each state and each transition. 
Unique IDs can be assigned to transitions based on concatenating the hashes of the connectivity graphs of the initial and final states obtained following the approach introduced in \cite{beland2011kinetic}.  
The hierarchical nature of the trajectory is clearly apparent: the same states are either frequently revisited in quick succession or not visited for extremely long times; a similar pattern is observed for transitions. 
In the figure, inter-super-state transitional periods are clearly apparent as changes of horizontal reoccurrence patterns, which are directly captured by the GPCCA analysis used to color each point. 
Also apparent in the figure is the presence of ``meta"-states composed of multiple super-states that the trajectory revisits in succession, a reflection of the hierarchical nature of the potential energy surface.
Super-states contain large numbers of states (e.g., many thousands for the teal state). 
Also apparent is the fact that new states can continually be observed within super-states even after multiple long visits. 
This is consistent with the extremely large number of disordered/frustrated configurations that characterize these systems.  
In contrast, some other super-states (e.g., pink) contain a very small number of states that are constantly revisited. 

While this type of visualization is insightful and enables a compression of the entire trajectory into a much smaller set of different coarse epochs that correspond to visits to different super-states, it fails to answer a number of important questions:

\begin{enumerate}
    \item What types of topological changes are associated with these transitions?
    \item How are different transitions related to each other? While identical transitions can be detected with the discrete graph labeling methodology, it is not possible to compare and classify them.
    \item How can we identify and characterize transitions that are the triggers of the long-lived topological changes?
\end{enumerate}

\subsection{Clustering Transitions}
\begin{figure}
    \centering
    \includegraphics[width=\linewidth]{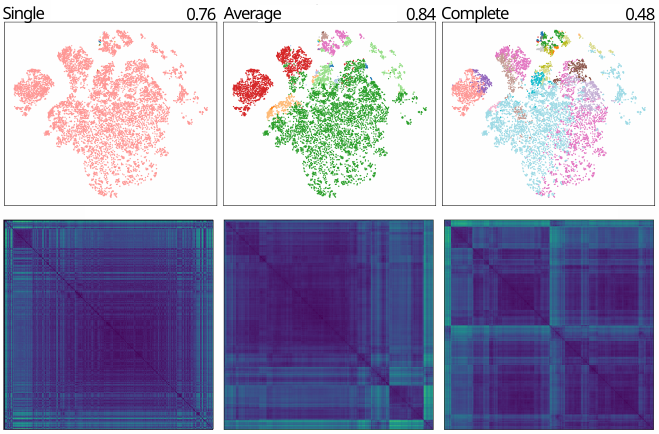}
    \caption{Top: A t-SNE~\cite{maaten2008visualizing} projection of the transition subset colored by the clusters formed by each linkage at $k=12$.
    While complete linkage seems to visually cluster better, it is important to note that t-SNE plots follow from complex non-linear mappings of the distance matrices that do not necessarily provide a complete and unbiased picture of the data. 
    Bottom: Corresponding distance matrices sorted by dendrogram leaf order. Cophenetic correlation coefficients for each clustering is reported in the title of each column.
}
    \label{fig:cluster_overview}
\end{figure}

Using the method described in \Cref{sec:methods}, we computed cross-scale graph-pooled Chebyshev signatures (c.f., \Cref{eq:signatures}) for each unique transition in our trajectory (as determined by the discrete transition IDs). 
\Cref{tab:hyperparams} enumerates the hyper-parameter values used for our experiment. 
Given that many transitions are collective in our system of interest, we opt for a long cutoff of 12 \AA; a shorter value could be more appropriate to favor spatial locality. 
Since our approach does not rely on spatial derivatives of the kernel, the choice of the order of the Wendland kernel is not critical. 
Finally, since normalized graph Laplacians have their spectrum bounded in $[0,2]$, the natural scale for the characteristic timescales of the heat kernels are of order unity. 
We therefore uniformly spread 4 probe times on a logarithmic scale between 1 and 10.  

\begin{table}
\centering
\begin{tabular}{|c|c|}
\hline
\textbf{Hyper-parameter} & \textbf{Value} \\
\hline
$r_c$                      & 12 \AA\\
$k_c$                      & 4                                    \\
$t$                        & $\{ 10^0, 10^{1/3}, 10^{2/3}, 10^1 \}$            \\
\textit{Wendland function} & $(1 - x)^{6}_{+}(35x^{2} + 18x + 3)$ \\ \hline
\end{tabular}
\caption{Hyper-parameters used to compute the cross-scale graph-pooled Chebyshev signatures.}
\label{tab:hyperparams}
\end{table}

\begin{figure}
    \centering
    \includegraphics[width=0.5\linewidth]{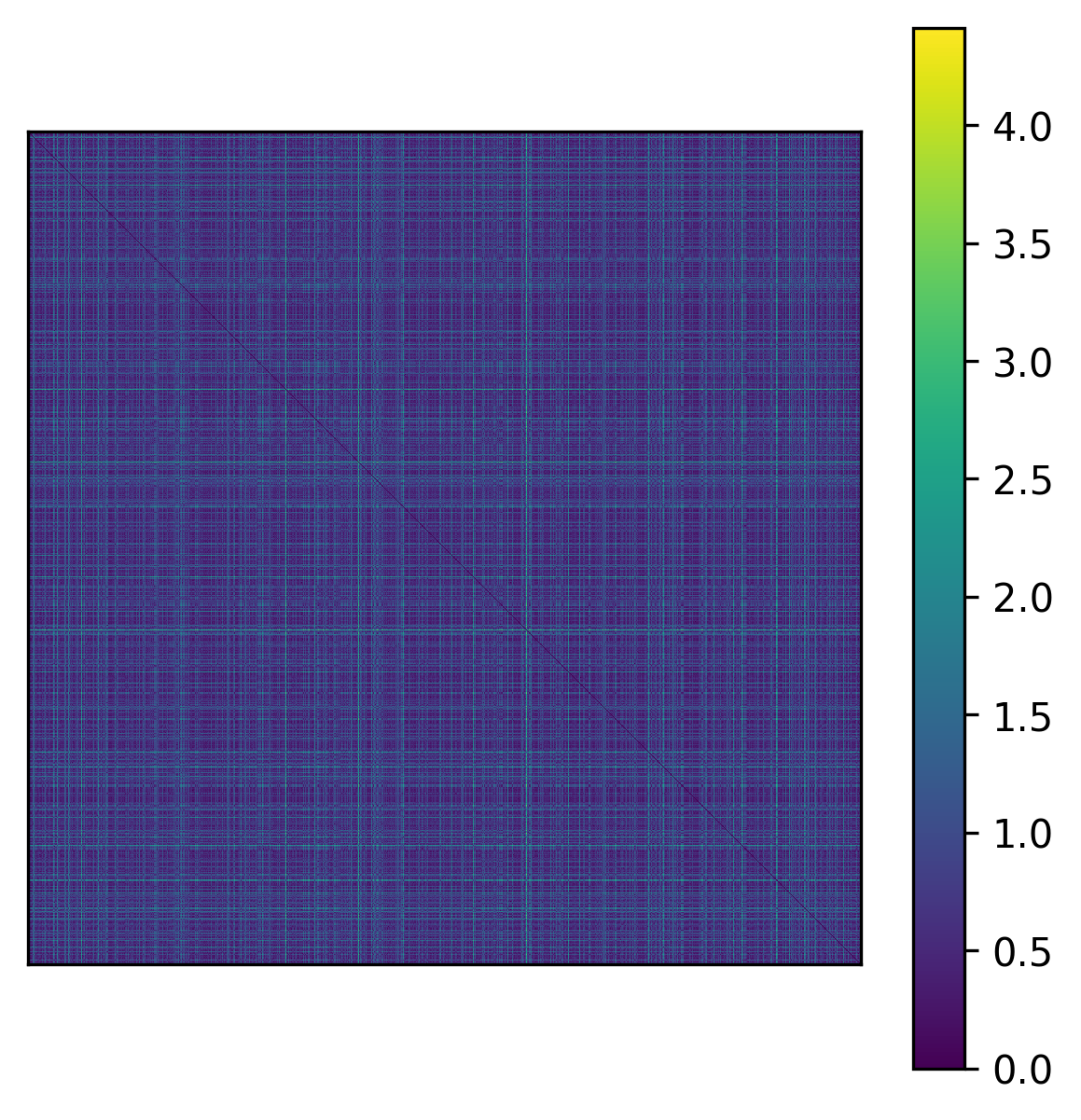}
    \caption{The unsorted distance matrix generated from taking the Wasserstein distance between the cross-scale energy signatures.}
    \label{fig:distance_matrix}
\end{figure}

We then defined a measure of distance between transitions by summing the respective Wasserstein metrics of each component of these signatures. The resulting matrix is reported in~\Cref{fig:distance_matrix}. The fact that the matrix is highly structured is immediately apparent: while certain transitions are clearly "unusual" as reflected from the fact that they are deemed dissimilar to most other transitions (green rows and columns), others are much more "typical", as many other transitions are deemed similar (blue). 

To make this statement more precise, the transition dataset was clustered using agglomerative hierarchical clustering (AHC)~\cite{murtagh2012algorithms}, a bottom-up method that begins with each data point in its own cluster and then iteratively merges the most similar pair of clusters according to a chosen distance/linkage rule until they all merge into a single cluster.  
As will be shown below, this manner of clustering facilitates interpretation as it naturally allows for a multi-resolution analysis in terms of parent and children clusters. 

Three linkages are typically used in AHC in conjunction with general distance definitions: single linkage merges cluster to minimize the increase in the smallest pairwise inter-cluster distances, complete linkage merges clusters to minimize the maximal pairwise inter-cluster distance, while average linkage merges clusters so as to minimize the increase in average pairwise inter-cluster distances. 
A fourth popular option, Ward linkage, is not appropriate here as our definition of distance does not correspond to a Euclidean embedding. 
The cophenetic correlation coefficient \cite{farris1969cophenetic}, a measure of the correlation between the pairwise distances and the "height" at which clusters merge, was then computed for all three approaches, a higher score indicating a clustering that is more faithful to the original distance definition. This metric, reported in \Cref{fig:cluster_overview}, clearly favors the average linkage over both single and complete for this application. Therefore, results from the average linkage will be reported below. 

As shown in the bottom of \Cref{fig:cluster_overview}, sorting the distance matrices according to the dendrogram leaf ordering clearly reveals distinct families of transitions, each corresponding to a blue diagonal block. 
The dendrogram leaves are sorted using the Bar-Joseph optimal leaf ordering algorithm~\cite{Bar-joseph.2001.FOLa}, which seeks to minimize the sum of distances between adjacent leaves.
The hiearachical nature of the clustering is also apparent, as strongly similar subsets within each family are also apparent.

In order to further illustrate the clustering, we rendered the entire data-set as a t-SNE~\cite{maaten2008visualizing} projection colored by the clusters formed by each type of linkage (\Cref{fig:cluster_overview}) at $k=12$. 
While these low-dimensional projections are difficult to interpret due to the non-linear nature of the mapping, it is apparent that average linkage appears to capture different isolated groups of transitions, while leaving a the large central population clustered together. 
In contrast, complete linkage partitions this central population into a number of groups. 
While it is difficult to judge the relevance of these differences, it is apparent from the sorted distance matrices using average linkage that transitions in this central population are in fact very similar to each other, but are being split apart under complete linkage. 
In contrast, single linkage clearly under-resolves the different groups.

\begin{table}[ht]
\centering
\begin{tabular}{rrl}
\hline
\textbf{Cluster ID} & \textbf{Cluster Size} & Description \\
\hline
2272  & 1 & Complete reversal of incomplete five-fold axis network\\
45900 & 51 & Nucleation of face-sharing icosahedral network\\
45931 & 128 &  Nucleation of face-sharing icosahedral network\\
45930 & 17962 &  Small amplitude localized surface changes in non-111 facets \\
45935 & 3767 & Small amplitude surface and volume changes around  non-111 facets\\
45922 & 14 &  Collective surface and volume reorganization around  non-111 facets\\
45926 & 27 & Rotation of face-sharing icosahedral network \\
13750 & 1 & Complex rearrangement of icosahedral network\\
45934 & 9851 &  Surface exchange rings\\
45847 & 4 & Stabilization five adjacent 111 facets\\
45925 & 28 & Stabilization five adjacent 111 facets\\
45933 & 14 & Stabilization five adjacent 111 facets \\
\hline
\end{tabular}
\caption{Summary of the clusters identified by the AHC algorithm with average linkage for 12 clusters.}
\label{tab:cluster_size}
\end{table}

\subsection{Mechanistic Interpretation of the Transition Families}
\begin{figure}
    \centering
    \includegraphics[width=\linewidth]{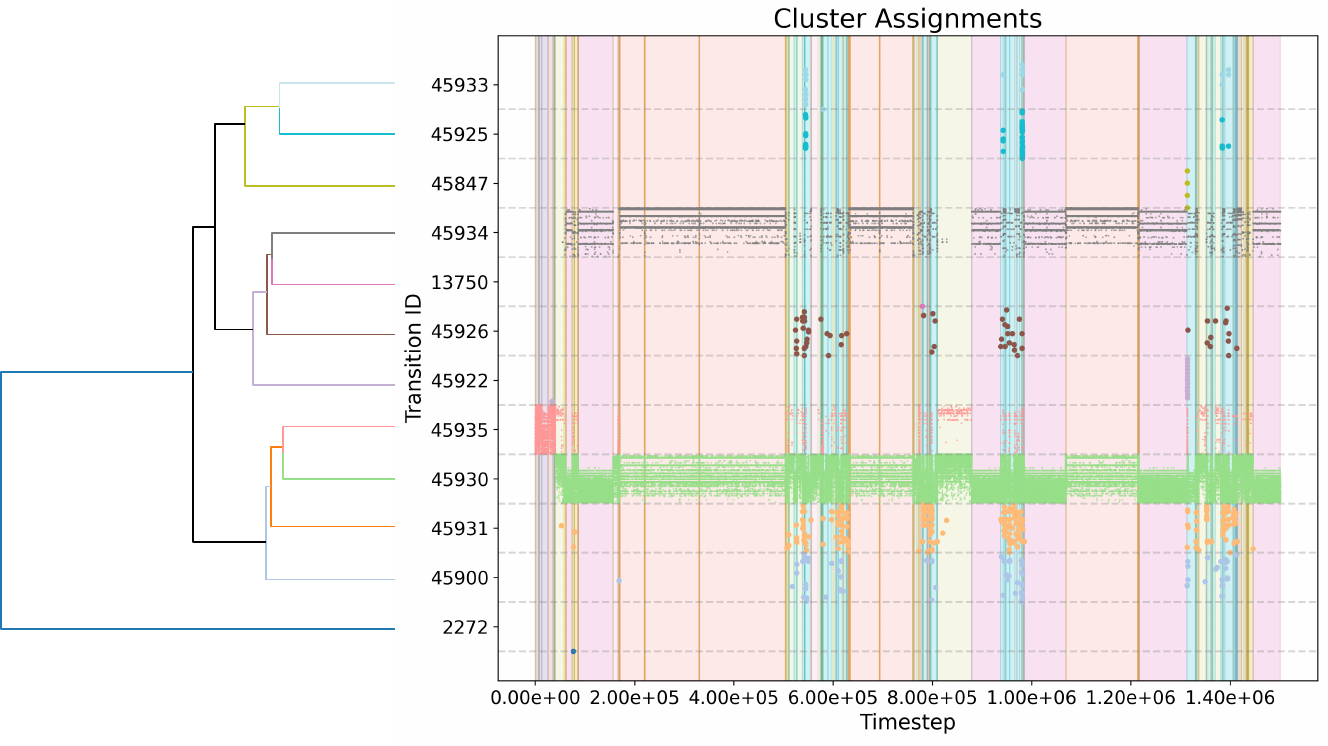}
    \caption{Left: Dendrogram of the AHC clustering using average linkage for $k=12$. The clusters are labeled as the agglomerative step at which they were merged into existence. Small labels correspond to very well separated clusters that were created early on in the agglomerative process. The index of the last merger corresponds to the total number of transitions. Right: the transition ID vs time plot for each cluster. The background is shaded according to the PCCA cluster the region is assigned to.
    }
    \label{fig:dendrogram}
\end{figure}

As the distance metric used to cluster transitions is very abstract, we assessed whether the identified clusters of transitions can be interpreted mechanistically in a way that allows for an intuitive understanding of the nature of the corresponding topological changes. 
To do so, we consider the clustering into $k=12$ families using the average linkage. 
The dendrogram corresponding to this clustering is reported in \Cref{fig:dendrogram}. 
In this figure, super-states identified by PCCA are shown as background colors and transitions are sorted according to their leaf order in the dendrogram, meaning that transitions from the same cluster appear near each other on the figure. The reported cluster labels correspond to the illustrated dendrogram.

We use two complementary approaches to form a mechanistic interpretation of these groups: displacement maps and common neighbor analysis (CNA) \cite{honeycutt1987molecular,faken1994systematic}. Displacement maps correspond to simple visualizations where the initial and final states of the particle are first translationally and rotationally aligned, and displacement vectors drawn from the initial to the final position of each atom. 
In parallel, we analyze the transition families in terms of a common-neighbor analysis; for a bond between two atoms $a$ and $b$, CNA associates a signature $ijk$ defined as 
\begin{itemize}
\item \textbf{i} — number of common neighbors shared by $a$ and $b$
    \item \textbf{j} — number of bonds among those common neighbors
    \item \textbf{k} — length of the longest contiguous chain among those bonds
\end{itemize}
In the following, we focus on 3 families of CNA bonds in order to identify recurring patterns within each cluster. 
555 bonds have 5 shared neighbors connected by a closed path, corresponding to a pentagonal arrangement perpendicular to the bond. 
These are key signatures of bulk icosahedral/decahedral order are found at the core of 5-fold symmetry axes. 
322 bonds are surface bonds commonly found at the edges joining adjacent (111) facets in icosahedral configurations. 
Finally, 311 bonds are surface bonds found in the plane of (111) facets. 
While other bonds types are potentially of interest (e.g., 421 indicating local FCC order or 422 indicating stacking faults), the fact that these nanoparticles were observed in previous studies to oscillate in-and-out of icosahedral configurations \cite{huang2018direct} motivates the choice of key surface and bulk icosahedral signatures. 

The results of a manual analysis and interpretation of these clusters in reported in \Cref{tab:cluster_size}. While the interpretation remains subjective, clear common patterns emerge within each transition cluster and families of clusters. 

Cluster 2272 contains a single transition. A CNA visualization reveals a collective and dramatic change in the icosahedral ordering within the cluster which can be described as a reversal or inversion of the icosahedral order where a group of 5 5-fold axis flips from pointing to the right in the initial state to pointing to the left in the final state. This is a telling example of the extreme complexity of the energy landscape of these systems where collective transitions can totally upend the very structure of the nanoparticle.

\begin{figure}
    \centering
    \includegraphics[width=0.3\linewidth]{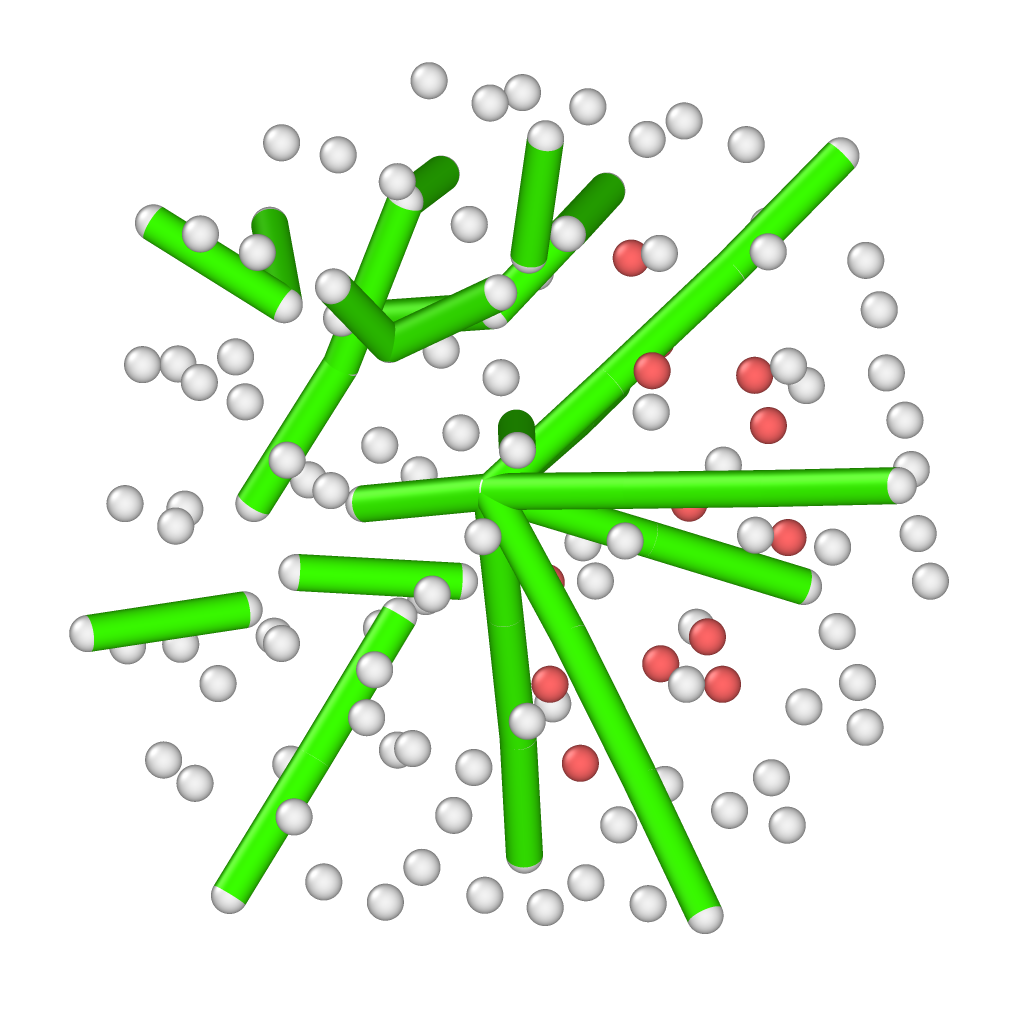}
    \includegraphics[width=0.3\linewidth]{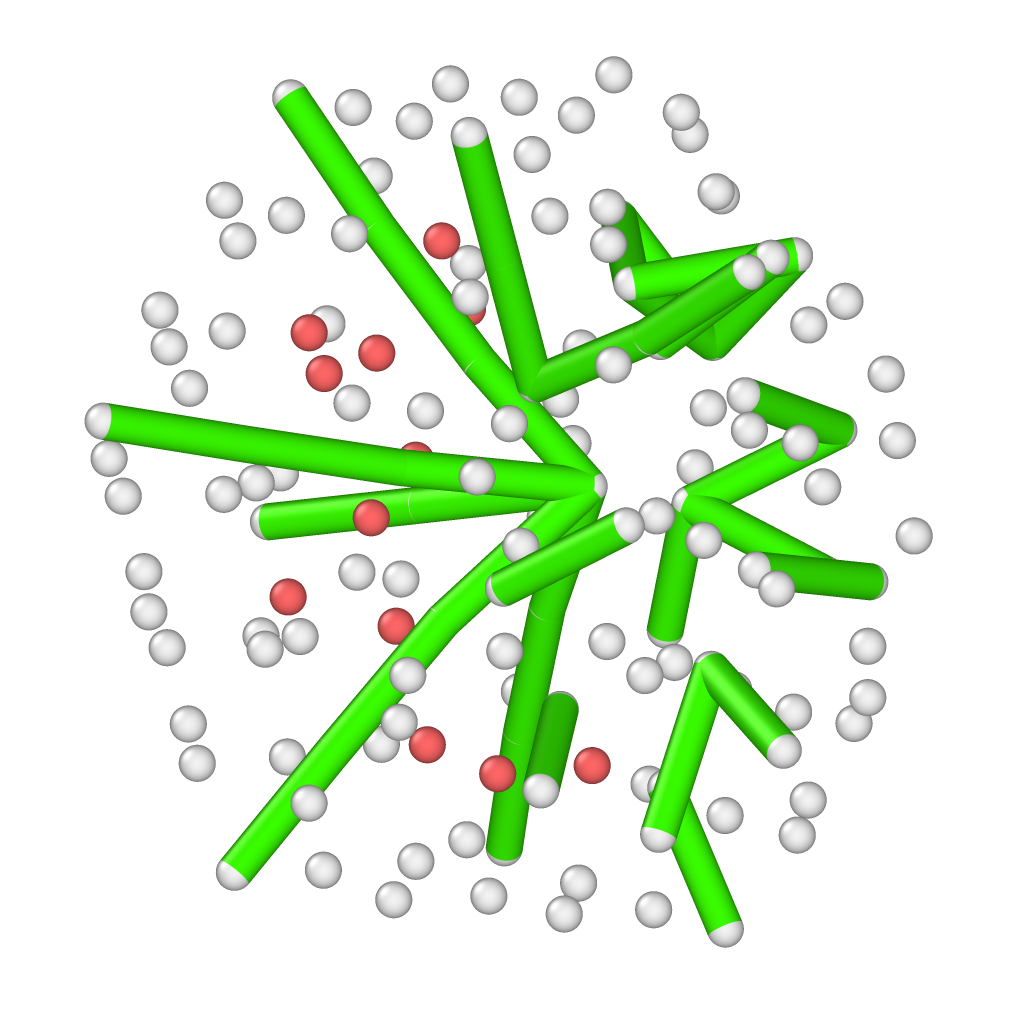}
    \caption{Illustration of the sole transition in the 2272 cluster. Atoms are color coded according to standard CNA definitions (red: FCC, white: other) and 555 bonds are shown. Top: initial state; Bottom: final state. The configurations are rotationally and translationally aligned.}
    \label{fig:2272}
\end{figure}
\begin{figure}
    \centering
    \includegraphics[width=0.3\linewidth]{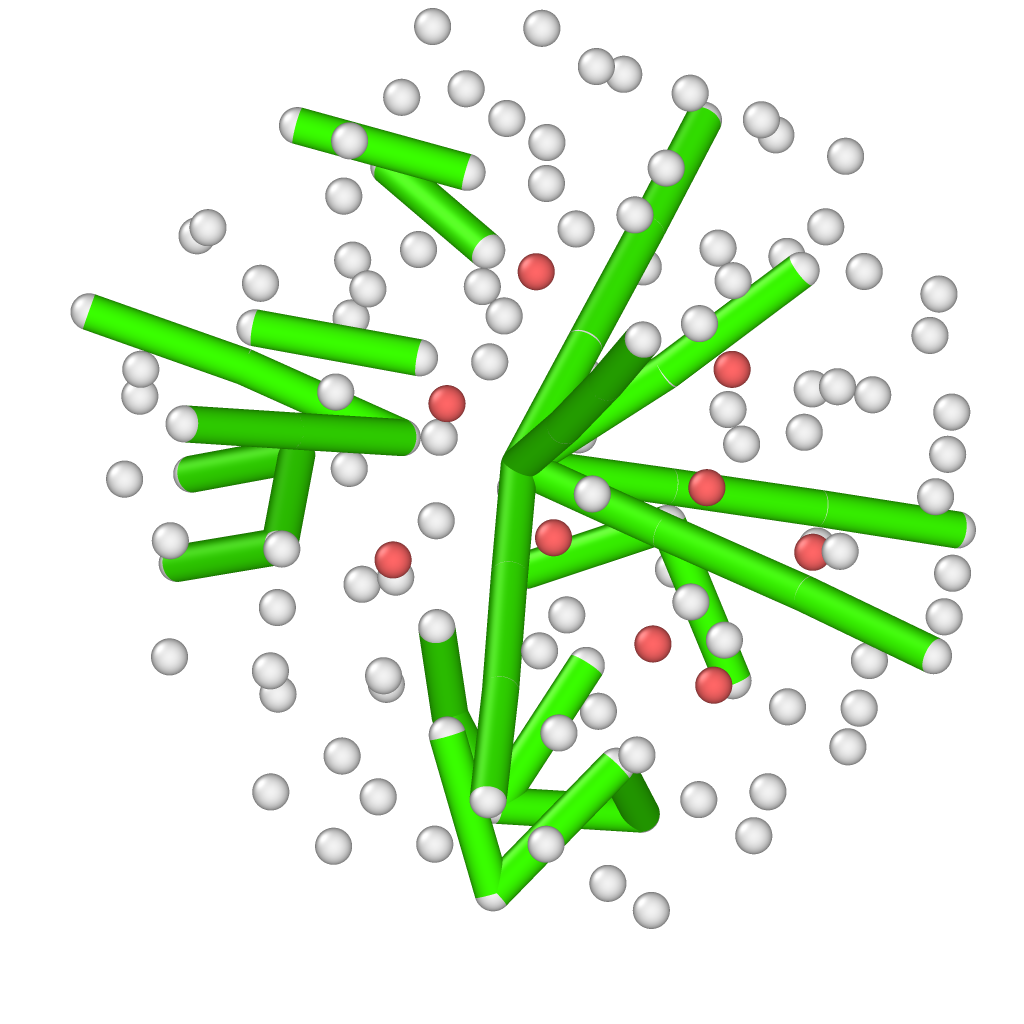}
    \includegraphics[width=0.3\linewidth]{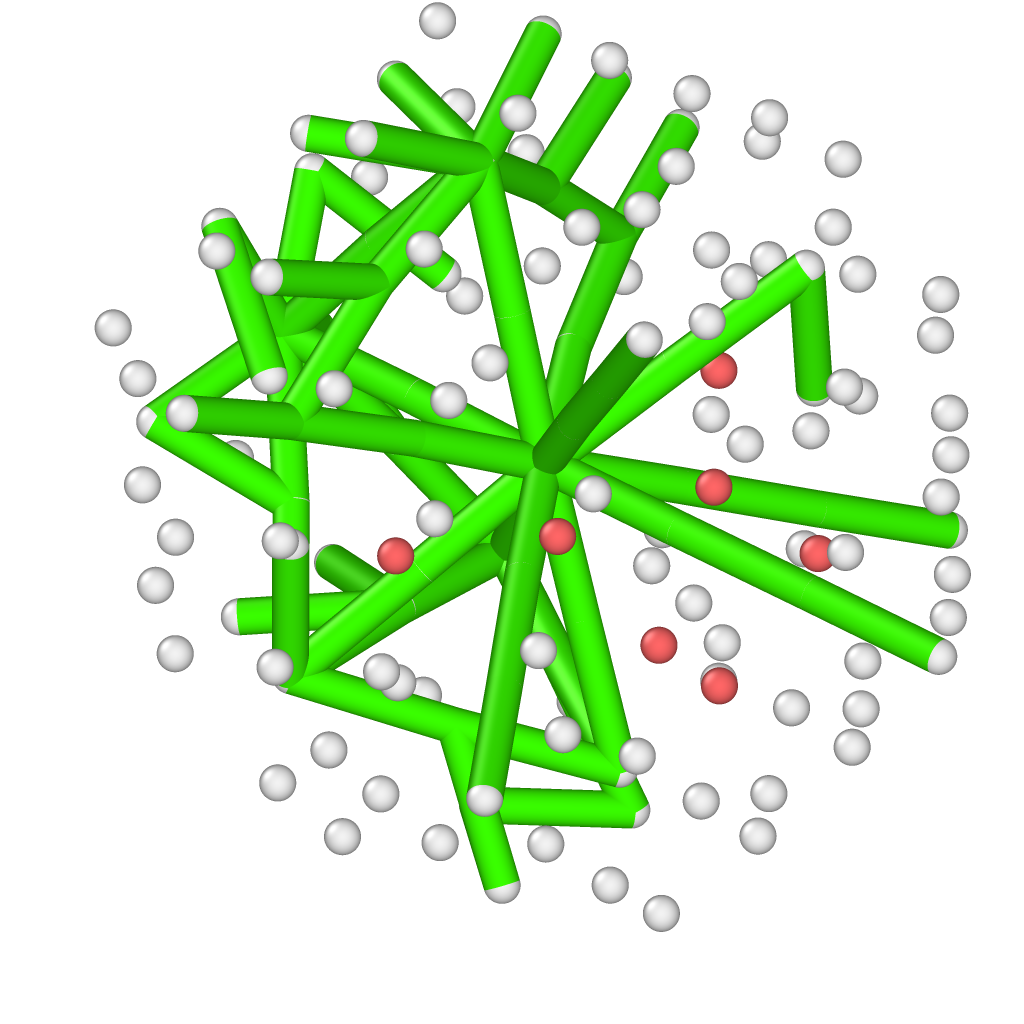}
    \caption{Illustration of a typical transition in the 45900 cluster. Atoms are color coded according to standard CNA definitions (red: FCC, white: other) and 555 bonds are shown. Top: initial state; Bottom: final state. The configurations are rotationally and translationally aligned.}
    \label{fig:49500}
\end{figure}

The second family contains two small clusters, 45900 and 45931. 
These contain configurations where the internal icosahedral network of the cluster evolves through the nucleation of a network of face-sharing tetrahedron defined by 555 bonds. 
This can be seen in \Cref{fig:49500} where a network of disordered 555 bonds in the left half of the particle --- a region where the "proper" icosahedral order is not established --- suddenly develops a regular structure through obtaining a face-sharing tetrahedron. This type of icosahedral order is also observed in bulk metallic glasses \cite{nelson1983order,nelson1989polytetrahedral}. 
We note that such networks are not present in the perfect icosahedral configuration where spokes of 555 bonds would emanate radially from the center of the particle, and therefore occur in regions where the icosahedral order is present but frustrated.

\begin{figure}
    \centering
    \includegraphics[width=0.3\linewidth]{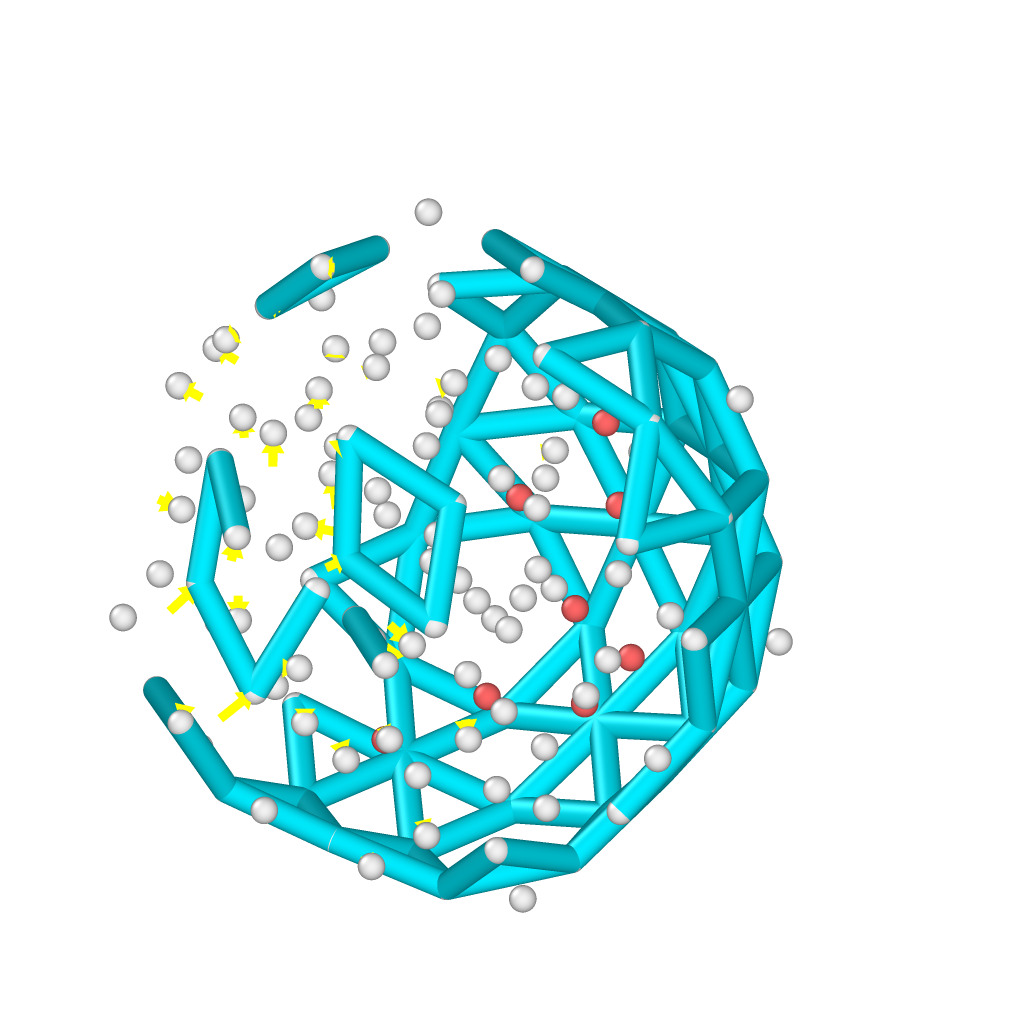}
    \caption{Illustration of a typical transition in the 45930 cluster. Atoms are color coded according to standard CNA definitions (red: FCC, white: other) and 311 bonds are shown. Yellow arrows correspond to the displacement between the initial and final state (final state shown)}.
    \label{fig:49530}
\end{figure}

The third family contains two extremely large clusters 45930 and 45935. 
These two clusters contain mostly transitions that affects the surface of the particle, although cluster 45935 appears to contain more displacements also affecting the bulk. \Cref{fig:49530} shows that these displacements are localized in regions where the surface does not exhibit (111) facets typical of icosahedral order. 
These unreconstructed surfaces are typically associated with nearby bulk regions where the icosahedral order is also not well established. 
The extremely large number of transitions observed in these clusters is consistent with the disordered/frustrated nature of these regions which creates a huge number of different local minima often connected by low barriers.

\begin{figure}
    \centering
    \includegraphics[width=0.3\linewidth]{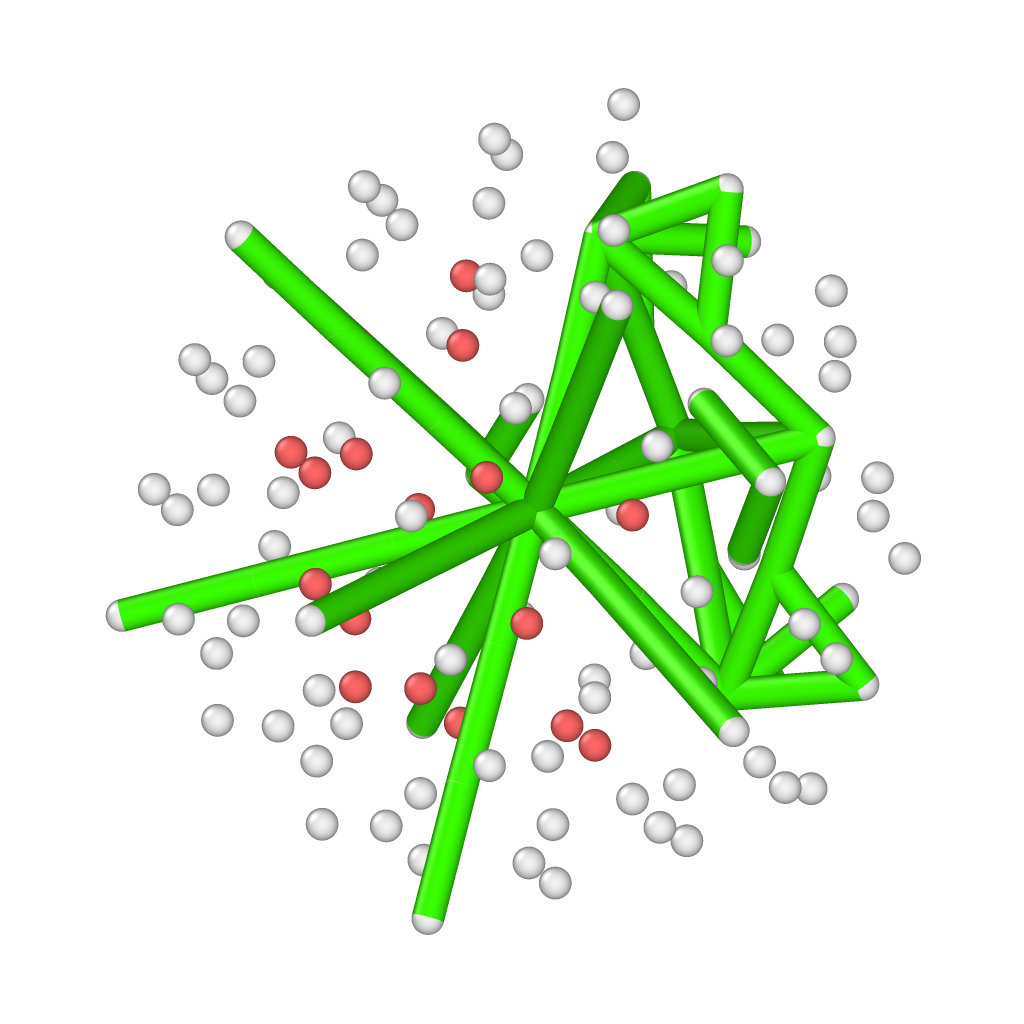}
    \includegraphics[width=0.3\linewidth]{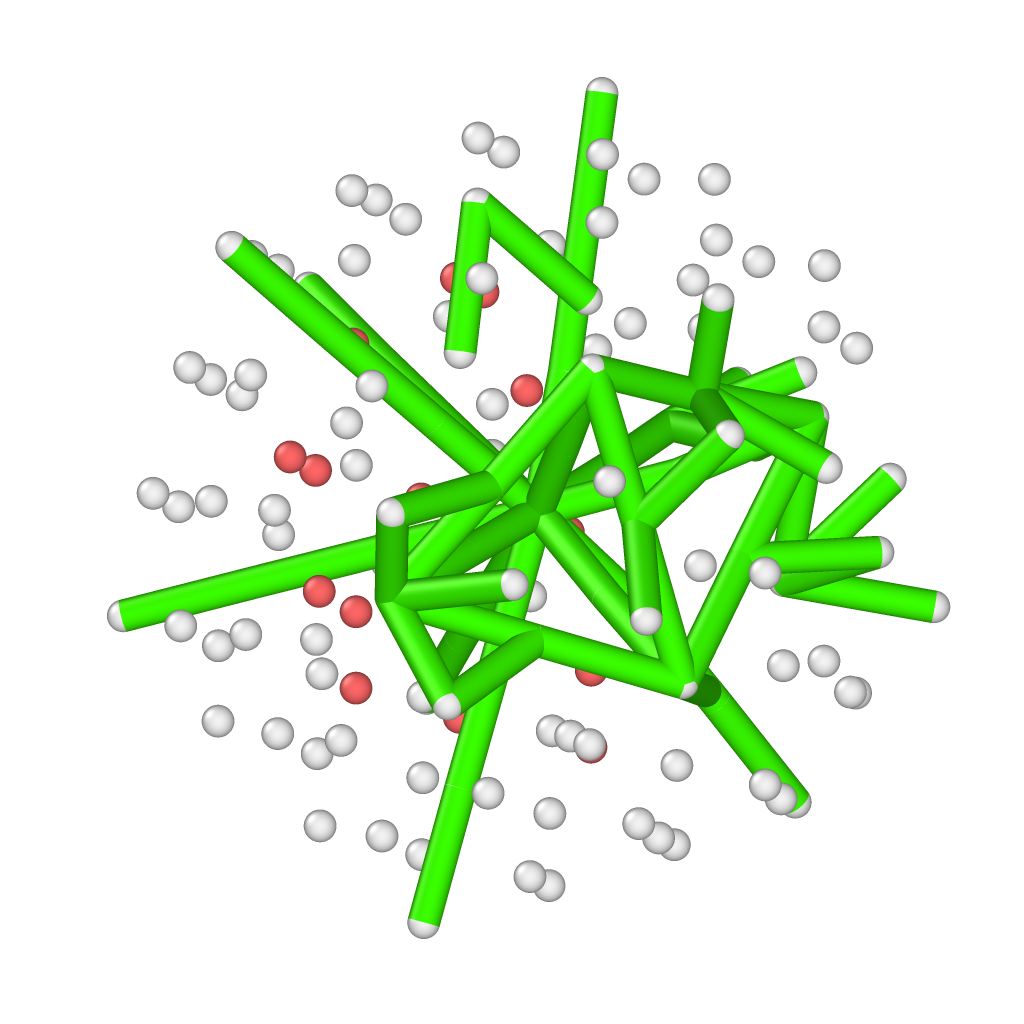}
    \caption{Illustration of a typical transition in the 45926 cluster. Atoms are color coded according to standard CNA definitions (red: FCC, white: other) and 555 bonds are shown. Top: initial state; Bottom: final state. The configurations are rotationally and translationally aligned.}
    \label{fig:45926}
\end{figure}

The fourth family contains clusters 45926 and 13750.  These are also associated with reorganization of the internal icosahedral bond network. 
For example, cluster 45926 contains transitions that can be described as rotations of face-sharing 555 tetrahedron, as shown in \Cref{fig:45926}.
These transitions act in regions where the icosahedral ordering is strong and spatially coordinated, but inconsistent with the rest of the nanoparticle that exhibits clear 555 radial spokes.

\begin{figure}
    \centering
    \includegraphics[width=0.3\linewidth]{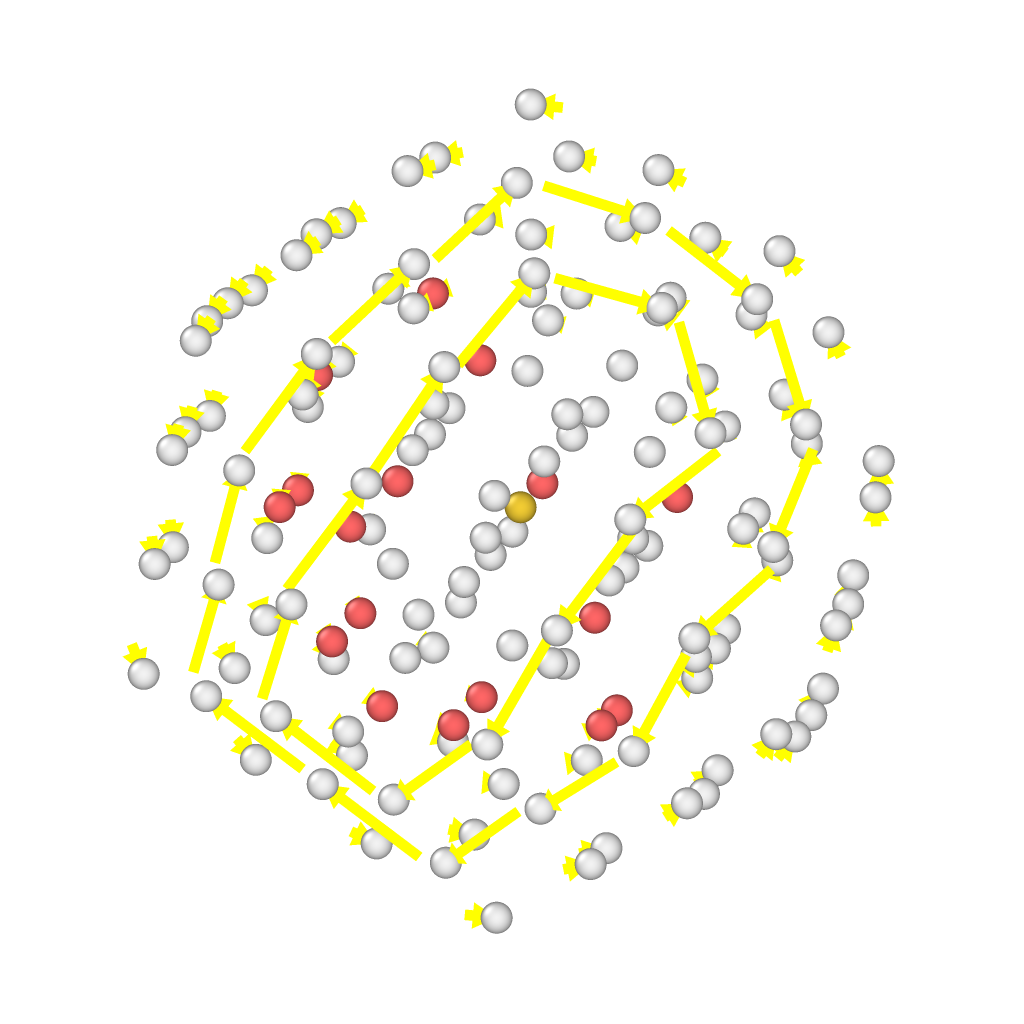}
    \caption{Illustration of a typical transition in the 45934 cluster. Atoms are color coded according to standard CNA definitions (red: FCC, white: other). Yellow arrows correspond to the displacement between the initial and final state (final state shown). The configurations are rotationally and translationally aligned.}
    \label{fig:45934}
\end{figure}

The fifth family contains the very large cluster 45934. A key characteristic of this cluster is the presence of surface transitions that exhibit various ring exchange patterns, where multiple closed groups of atoms exchange position with each other.  These rings take many shapes and sizes some very ordered, as the one shown in \Cref{fig:45934}, and some more complex. These transition typically have a limited impact on the bulk structure of the nanoparticle.

\begin{figure}
    \centering
    \includegraphics[width=0.3\linewidth]{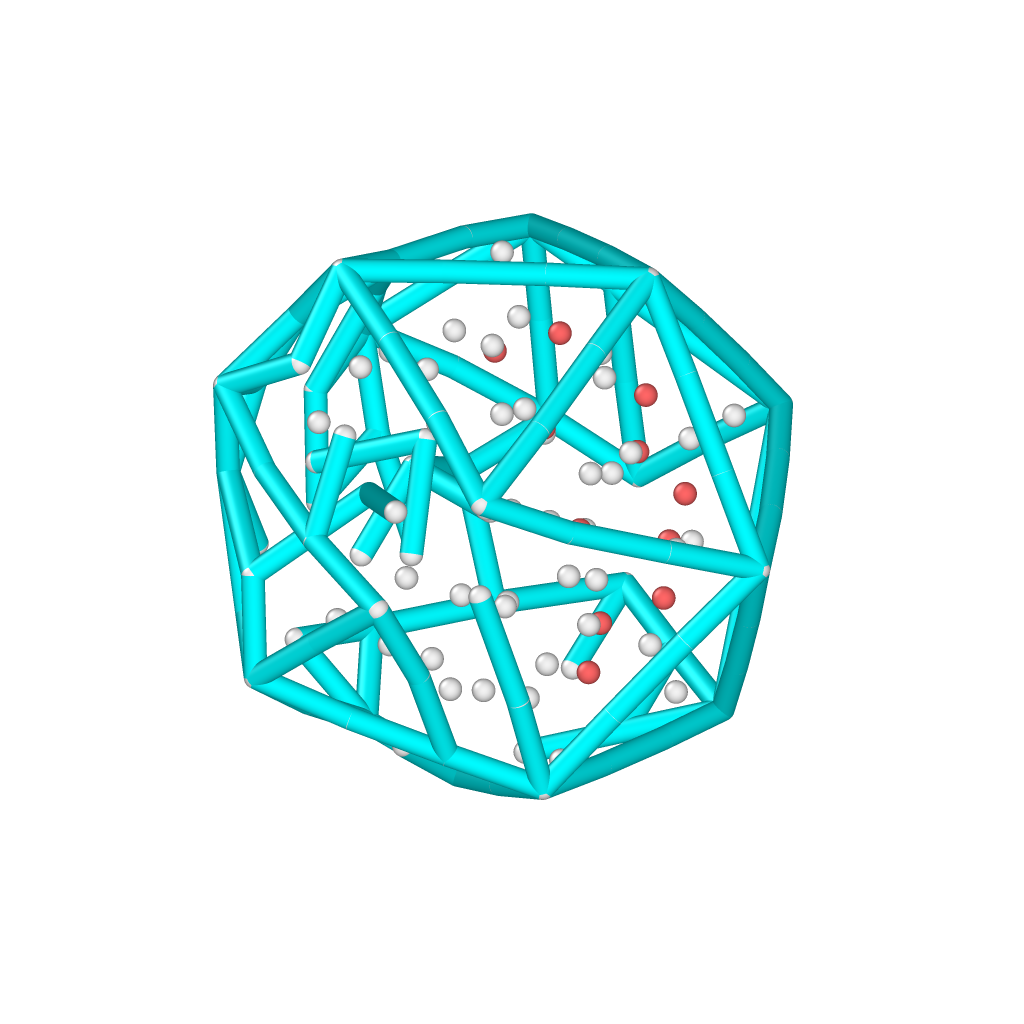}
    \includegraphics[width=0.3\linewidth]{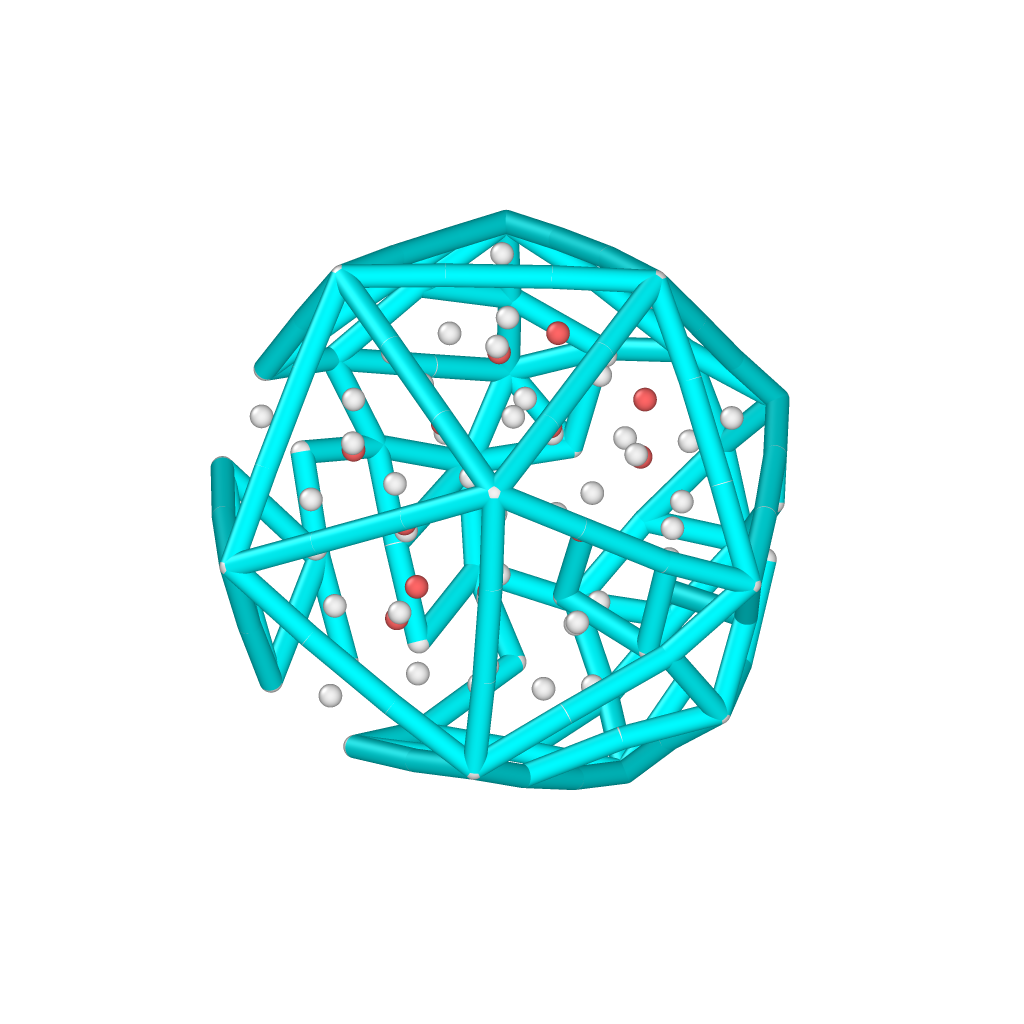}
    \caption{Illustration of a typical transition in the 45847 cluster. Atoms are color coded according to standard CNA definitions (red: FCC, white: other) and 322 bonds are shown. Top: initial state; Bottom: final state. The configurations are rotationally and translationally aligned.
}    
\label{fig:45847}
\end{figure}

The final family of clusters contains 3 small clusters: 45847, 45925, and 45933. 
These also contain mostly surface-localized transitions, although very different in nature. 
In this group, disordered surface regions collectively organize into groups of 5 (111) facets sharing a common vertex, forming a stable icosahedral cap from an initially disordered region on the surface. This type of transition is illustrated in \Cref{fig:45847} using the network of 322 bonds that are found at the intersection of the (111) facets. 

While difficult to quantify, the exercise of manually analyzing of these transitions has demonstrated that simple summary statistics such as net changes in CNA index counts or even per type transition summaries (e.g., how many type $ijk$ bond transformed into type $lmn$ bonds, etc.) are unable to capture the topological complexity of the transitions occurring in these systems to a remarkable extent, presumably because spatial correlations between these changes are central to identifying similarities. 
This is perhaps not surprising as the cross-scale Chebyshev signature was explicitly designed to capture such spatial correlations. 
Instead, this exercise required time-consuming manual analysis with OVITO \cite{Stukowski.2009.VAA}, a powerful visualization tool, to identify the common features shared by members of each cluster. 
However, this exercise has also convinced us that clustering based on cross-scale Chebyshev signatures naturally captures these non-trivial patterns. Therefore, while interpreting the clusters still requires some human effort, manually classifying the type of complex transitions observed here would have been impossible using either simple summary statistics or visualization tools alone.

As shown in \Cref{tab:cluster_size}, the mechanistic interpretation of these families is remarkably consistent with the dendrogram's structure, as manual annotation is directly reflected in the parent structure of the different groups. 
This indicates that hierarchical clustering captures coarse similarities between groups of similar transitions. 

\subsection{Trajectory Analysis}
\begin{figure}
    \centering
    \includegraphics[width=\linewidth]{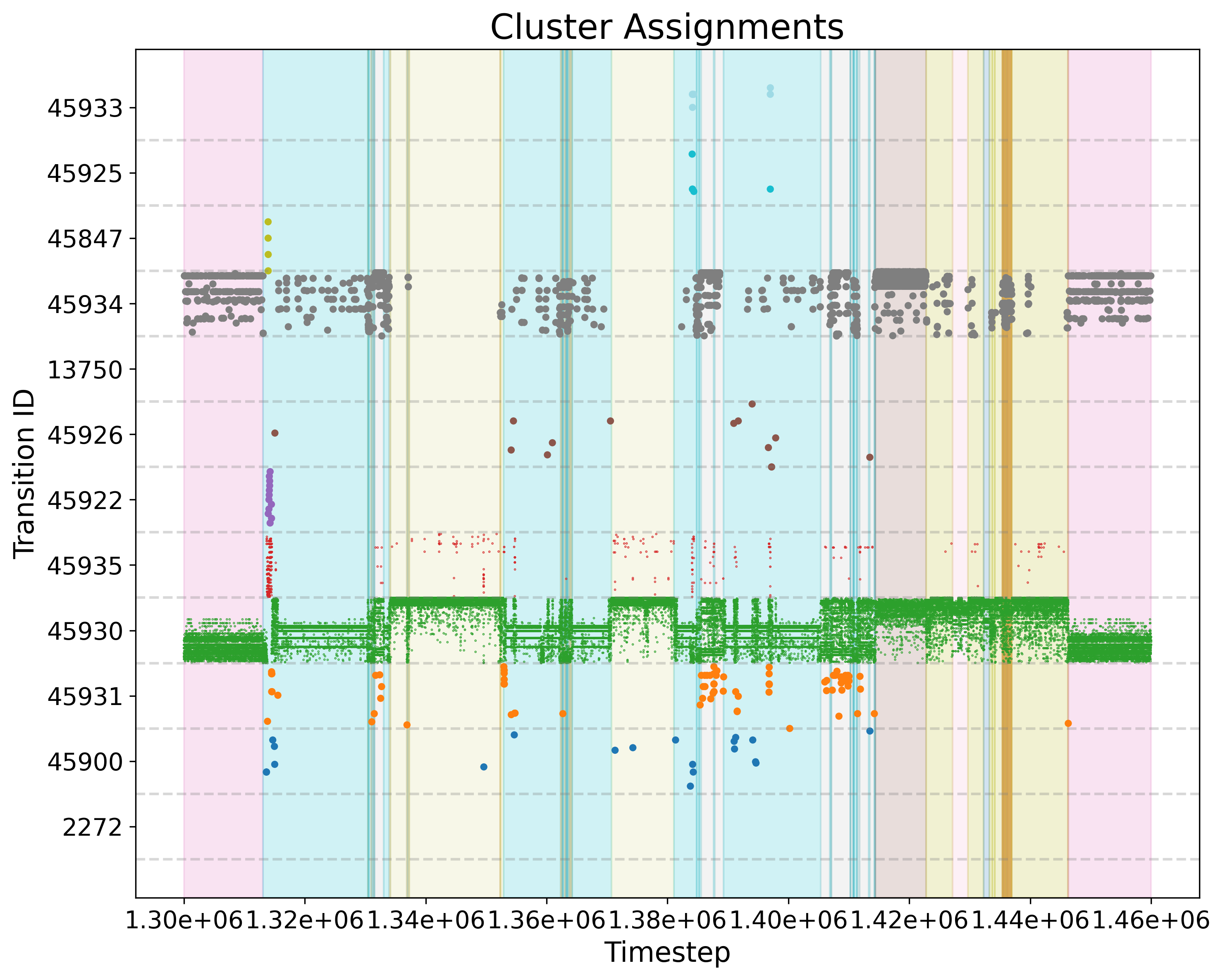}
    \caption{In-depth analysis of a section of the full trajectory. The transitions are sorted according to their dendrogram leaf order. Each transition cluster is rescaled to occupy the same fraction of the horizontal axis. The background colors correspond to super-states identified using the PCCA algorithm.}
    \label{fig:zoom_trajectory}
\end{figure}
We now demonstrate how the transition clustering dramatically facilitates the analysis of long ParSplice trajectories. 
For ease of visualization, we focus on a subset of whole the trajectory where a number of super-state to super-state transitional regions punctuate long periods of repetitive transitions. 
The corresponding sequence is reported in \Cref{fig:zoom_trajectory}. 
In this figure, vertical colored regions correspond to super-states identified by the GPCCA algorithm.  
Transitions were here sorted according to their respective dendrogram leaf order so that transitions that are part of the same cluster are mapped to a similar vertical position and rescaled so that each cluster occupies the same height in the figure.  

The figure shows that each super-state possesses a unique fingerprint in terms of exactly which transitions occur.
First, different super-states clearly exhibit different distributions of transitions across clusters. For example, transitions from the 45934 cluster occur extremely frequently, but but the pale yellow super-state around transition 1,340,000 exhibits none. Similarly, transitions in cluster 45926 only occur in the blue super-state. 
In contrast, transitions in cluster 45930, corresponding to small amplitude surface reconstructions away from (111) regions systematically occur in all super-states. 
The distribution of transition clusters is therefore a strong signature of each super-state. 

More importantly, the occurrence of transitions within small clusters, corresponding to comparatively rare transitions, often directly correlates with transitional regions between super-states.
The most striking example is the pink-to-blue transitional region around transition 1,314,000. 
In a short amount of time, a large number of comparatively rare transitions were observed in rapid sequence (as captured by quasi-vertical pattern of transitions from small clusters). 
This pattern repeats across many of the transitional regions where rare transitions are over-represented. 
This demonstrates two important points: first transitional regions between super-states are characterized by distinct topological reorganizations that directly relate to the nucleation and reorganization of internal icosahedral order, while intra-super-state regions consist in vast majority of small amplitude surface and volume reconstructions in regions where the icosahedral order is lacking or defective. 
Second, transitional regions are not always sharp. For example, while the transitional region around transition 1,445,000 appears to have been triggered by a single event in cluster 45931, other transitional regions, e.g., around transition 1,385,000, experience multiple transitions from the rare groups in succession. 
This is a signature of a frustrated process where  multiple reorganization attempts combining surface reconstructions and nucleation or reorganizations of the internal icosahedral network occur before the system stabilizes into a new super-state. 
This is broadly consistent with the understanding of these systems where previous studies have shown that stable changes in topology typically result from complex multi-step processes \cite{huang2018direct}.
The transition clustering directly highlights these important structural transitions and enables their classification.

%% file: 03_discussion.tex
\section{Discussion}

\label{sec:discussion}
This study exemplifies the potential benefits of automated transition analysis and clustering in situations where the intrinsic complexity of the physical system and the sheer volume of data makes traditional manual visualization and analysis extremely time-consuming and burdensome. 
Instead, by leveraging abstract representations of the system in terms of invariants of transition operators that connect representations of the system before and after each transition, it becomes possible to directly compare and classify extremely large numbers of transitions in a way that makes physical patterns naturally emerge. 
As such, these methods directly complement and augment the arsenal of trajectory analysis techniques. 

For example, while the use of conventional local order parameters such as CNA was instrumental to interpret the families of transitions that were identified, the present exercise shows that \textit{classifying} these transitions on the basis on these order parameters is inefficient. 
Indeed, summary statistics of CNA changes along individual transitions have proven woefully insufficient at capturing and taming the complexity of the system under consideration for at least two reasons. 
First, most transitions in these systems are extremely complex and collective; as large numbers of CNA bonds of different nature are simultaneously created and destroyed. 
Furthermore, each family of transition contains considerable internal diversity due to the extremely large number of topological variants possible in frustrated systems. 
Second, CNA analysis require the definition of discrete bonds. 
For off-lattice systems with disorder like the one investigated here, bonds can be created or destroyed on the basis of extremely small displacements, which greatly complicates and obscures the analysis even when using energy-minimized configurations. 
We found that defining bonds based on a fixed cutoff radius was hopelessly noisy, and that while advanced techniques like the solid-angle based nearest-neighbor algorithm (SANN), \cite{van2012parameter} dramatically improved the situation, it remained extremely difficult to quantitatively extract robust patterns. 
However, the use of SANN was instrumental in enabling the classification of the different clusters, as clear visual patterns nonetheless emerge.
Finally, we observed that a large number of transitions are globally almost ``neutral" with respect to changes in CNA populations, in the sense that aggregate statistics of net changes show very little systematics even within each cluster. 
This can be explained by the fact that the before and after states of each transitions are often topologically quite similar, consistent with the observation that most individual transitions, while impacting a large fraction of the atoms in the system, do not materially change the topology of the nanoparticle.
Instead, sequences of multiple transitions are necessary to affect persistent topological changes. 
Overall, these observations are consistent with a previous study that shows that CNA characteristics of different PCCA clusters significantly overlap \cite{Huang.2017.CAA}, indicating that these representations capture different types of information about the trajectory. 
In contrast, trajectory clustering based on cross-scale energy signatures strongly correlate with GPCCA super-states, suggesting that capturing spatial correlations between local topological changes is essential to paint a complete picture of the trajectory.

Furthermore, while kinetic clustering such as GPCCA is clearly able to narrow down the location of transitional regions, they provide no indication of the type of transitions that trigger this change, nor of how these transitions are related to other transitional regions.
Furthermore, the actual transitional regions are diffuse, as multiple ``relevant'' transitions, often interspersed with ``uninteresting'' transitions, occur over a finite time, a feature that PCCA typically fails to capture (e.g., the transitional region around 1,315,000 in \Cref{fig:zoom_trajectory}).

While not explored in the present study, the fact that the cross-scale energy signatures clearly capture similarities between complex transitions indicate that they could enable a broad range of machine learning approaches that directly target transitions instead of individual states, e.g., energy barrier prediction. 
This fills a need in the atomistic ML toolkit, as methodologies for state characterization abound due to their direct use in designing interatomic potentials. 
The present study suggests how some of these state-wise methods could be generalized to transition-based analysis. 

One major limitation of the proposed approach is its computational scalability. 
Processing the data-set used in \Cref{sec:results} (20,000 transitions) took approximately a day and used 95GB of RAM at its peak; a smaller subset of 4,000 transitions took 4 hours and 20GB of RAM. 
The computation scales with the hyperparameters used as well as the size of the system, since high values of $k_c$ or many values of $\tau$ create $k_c+|t|$ matrices for each transition that are then compared using the Wasserstein distance ($O(n^3logn)$ per matrix). 
As such, finding a way to reduce the amount of matrices used in the computation or exploring an alternative distance metric to use in place of the Wasserstein distance is a promising direction for future work. 

\section{Conclusion}
A new method for the automated analysis and classification of complex trajectories observed in atomistic simulations was proposed. 
Using long-timescale simulations of small metallic nanoparticles as a prototypical example, we have shown that cross-scale energy signatures of transition matrices between Coulomb-matrix-like representations of the system remarkably capture the essence of the different transition families. 
This was demonstrated by analyzing the automated clustering obtained using a distance measure defined over transition signatures, which was used to classify almost 40,000 different transitions into physically meaningful families. 
An analysis of the trajectory in terms of these transition clusters revealed that rare transitions affecting either the internal icosahedral order of the particle or the surface reconstruction in terms of low-energy (111) facets dominates the slow evolution of the system, while large numbers of fast and repetitive transitions mostly affecting disordered surface and bulk regions of the particle occur in long visits to distinct super-states that occupy the majority of the simulation, consistent with previous analyses of these systems. 
By adopting a transition-centric view of the trajectory, this approach can dramatically simplify and automate the analysis of very large trajectories, complementing existing configuration or kinetics based approaches. 

\section{Acknowledgments}
DP acknowledges support from the DOE Office of Advanced Scientific Computing Research Competitive Portfolios program. Los Alamos National Laboratory, operated by Triad National Security, LLC, for the National Nuclear Security Administration of the U.S. Department of Energy (Contract No. 89233218CNA000001).